\documentclass[conference,10pt]{IEEEtran}
\IEEEoverridecommandlockouts

\usepackage{cite}
\usepackage{amsmath,amssymb,amsfonts,upgreek}
\usepackage{algorithmic}
\usepackage{textcomp}
\usepackage{xcolor}
\usepackage{cite}
\usepackage{url}
\usepackage{amsfonts}
\usepackage{amssymb}
\usepackage{mathrsfs}
\usepackage{euscript}
\usepackage{graphicx}
\usepackage{multirow}
\usepackage{algorithm}
\usepackage{algorithmic}
\usepackage{changepage}
\usepackage{color}
\usepackage{caption}
\usepackage{subcaption}
\usepackage{todonotes}
\usepackage{scalerel}
\usepackage{tikz}
\usetikzlibrary{svg.path}
\usepackage{balance}
\usepackage{svg}
\usetikzlibrary{fit, positioning, shapes.geometric, fit, arrows.meta, calc, backgrounds}
\usepackage{pgfplots}
\pgfplotsset{compat=newest}
 
\definecolor{darkgreen}{rgb}{0, 0.5, 0} 
\definecolor{lightpurple}{rgb}{0.7, 0.4, 1} 

\definecolor{orcidlogocol}{HTML}{A6CE39}
\tikzset{
orcidlogo/.pic={
\fill[orcidlogocol] svg{M256,128c0,70.7-57.3,128-128,128C57.3,256,0,198.7,0,128C0,57.3,57.3,0,128,0C198.7,0,256,57.3,256,128z};
\fill[white] svg{M86.3,186.2H70.9V79.1h15.4v48.4V186.2z}
svg{M108.9,79.1h41.6c39.6,0,57,28.3,57,53.6c0,27.5-21.5,53.6-56.8,53.6h-41.8V79.1z M124.3,172.4h24.5c34.9,0,42.9-26.5,42.9-39.7c0-21.5-13.7-39.7-43.7-39.7h-23.7V172.4z}
svg{M88.7,56.8c0,5.5-4.5,10.1-10.1,10.1c-5.6,0-10.1-4.6-10.1-10.1c0-5.6,4.5-10.1,10.1-10.1C84.2,46.7,88.7,51.3,88.7,56.8z};
}
}
\newcommand\orcidicon[1]{\href{https://orcid.org/#1}{\mbox{\scalerel*{
\begin{tikzpicture}[yscale=-1,transform shape]
\pic{orcidlogo};
\end{tikzpicture}
}{|}}}}

\usepackage[colorlinks=true,linkcolor=blue,citecolor=blue]{hyperref}
\newtheorem{proof}{Proof}
\newtheorem{proposition}{Proposition}

\def\BibTeX{{\rm B\kern-.05em{\sc i\kern-.025em b}\kern-.08em
T\kern-.1667em\lower.7ex\hbox{E}\kern-.125emX}}
\usepackage[most]{tcolorbox}

\newcommand*{\QED}[1][$\blacksquare$]{%
\leavevmode\unskip\penalty9999 \hbox{}\nobreak\hfill
\quad\hbox{#1}%
}

\begin{document} 
\author{
$
\text{Jalal Jalali}^{\orcidicon{0000-0002-3609-6775}}\ \IEEEmembership{Member, IEEE}, 
\text{Rodrigo C. de Lamare}^{\orcidicon{0000-0003-2322-6451}}\ \IEEEmembership{Senior Member, IEEE}
$.
}

\author{
\IEEEauthorblockN{
~Jalal Jalali\IEEEauthorrefmark{2},
~Mostafa Darabi\IEEEauthorrefmark{4}
\IEEEauthorrefmark{5},
and
Rodrigo C. de Lamare\IEEEauthorrefmark{2}\IEEEauthorrefmark{3}}
\IEEEauthorblockA{\IEEEauthorrefmark{2}A Centre for Telecommunications Studies, Pontifical Catholic University of Rio de Janeiro, Brazil}
\IEEEauthorblockA{\IEEEauthorrefmark{4}Department of ECE, University of British Columbia, Vancouver, BC V6T 1Z4, Canada\\}
\IEEEauthorblockA{\IEEEauthorrefmark{5}Wireless Communication Research Group, JuliaSpace LLC, Chicago, IL, USA\\}
\IEEEauthorblockA{\IEEEauthorrefmark{3} University of York, United Kingdom\\}
Emails: 
\href{mailto:josh@juliaspace.com}
{\texttt{josh@juliaspace.com}},
\href{mailto:mostafadarabi@ece.ubc.ca}
{\texttt{mostafadarabi@ece.ubc.ca}},
\href{mailto:delamare@puc-rio.br}
{\texttt{delamare@puc-rio.br}}
\vspace{-4mm}
}

\title{\huge
RHOSI:
Efficient Anti-Jamming Resource Allocation with Holographic Surfaces in UAV-enabled ISAC}
\maketitle

\begin{abstract}
This paper investigates the susceptibility of Integrated Sensing and Communication (ISAC) systems to hostile jamming, focusing on an aerial Reconfigurable Holographic Surface (RHS)-aided unmanned aerial vehicle (UAV). The proposed framework, termed RHOSI, enhances ISAC’s resilience by dynamically shaping the wireless propagation environment. Specifically, RHOSI introduces a strategy to improve jamming resistance by jointly optimizing transmit beamforming at the hybrid base station, RHS phase shift configuration, and UAV spatial deployment, while ensuring the required echo signal-to-interference-plus-noise ratios for reliable sensing. The resulting non-linear optimization problem features highly coupled variables, which are decomposed into sub-problems and solved using an alternating optimization (AO) approach. Simulation results confirm the practicality and effectiveness of RHOSI in significantly improving the throughput and robustness of ISAC under adversarial jamming.

\end{abstract}

\section{Introduction}
\indent 
\IEEEPARstart{T}{he} evolution of wireless networks forecasts sixth generation (6G) systems as intelligent platforms supporting applications such as autonomous vehicles, telemedicine, and intelligent manufacturing~\cite{10464825}. Integrated sensing and communication (ISAC) has emerged as a key enabler, enhancing spectrum efficiency by jointly exploiting infrastructure for both functions~\cite{9737357,8288677,9124713,10436573}. Yet, terrestrial ISAC systems are vulnerable to line-of-sight (LoS) blockages, limiting sensing and communication reliability.  

Reconfigurable holographic surfaces (RHS) are viewed as a core 6G technology for the next-generation wireless networks~\cite{250321542}. In this context, RHS-assisted ISAC has been shown to improve performance by joint beamforming at the base station (BS) and RHS:~\cite{9264225} enhanced radar detection probability while meeting SINR constraints,~\cite{9364358} maximized radar SNR while ensuring communication SNR, and~\cite{10042240} proposed iterative BS–RHS–user beamforming to maximize sum-rate under sensing constraints. These works confirm the effectiveness of RHS in balancing sensing and communication.  

Unmanned aerial vehicles (UAVs), with low cost and mobility, enable flexible RHS deployment to bypass obstacles~\cite{10456885,8470897,11200993}. UAV-mounted RHS can create LoS links and boost received signals via passive beamforming~\cite{9599592}. In~\cite{9424472}, aerial RHS placement and phase shifts were optimized to resist jamming, while~\cite{9810528} jointly optimized UAV trajectory, RHS, and power budgets to maximize rate under jamming. Aerial RHS also improves sensing accuracy when BS–target links are obstructed, reducing blockage probability and enabling directional beamforming~\cite{9729746}.  

In this work, we study a novel scheme where a UAV is empowered by RHS to mitigate malicious jamming in ISAC. Our objective is to minimize network power consumption through joint optimization of BS active beamforming, RHS passive beamforming, and UAV positioning under echo signal-to-interference-plus-noise ratio (SINR), BS power, RHS coefficients, and flight constraints. This leads to a non-convex mixed-integer nonlinear program (MINLP). To solve it, we develop an alternating optimization (AO)-based algorithm, termed \textbf{RHOSI} (\underline{R}esource allocation for anti-jamming \underline{HO}lographic \underline{S}urface UAV-enabled \underline{I}SAC), leveraging semi-definite relaxation, the big-M method, and successive convex approximation for efficient suboptimal solutions.

The rest of this paper is structured as follows. Section II introduces the system model and the problem formulation. Section III details the proposed RHOSI algorithm, whereas Section IV presents and discusses the simulation results. Section V draws the conclusions.

\begin{figure}
\centering
\includegraphics[width=0.425\textwidth]{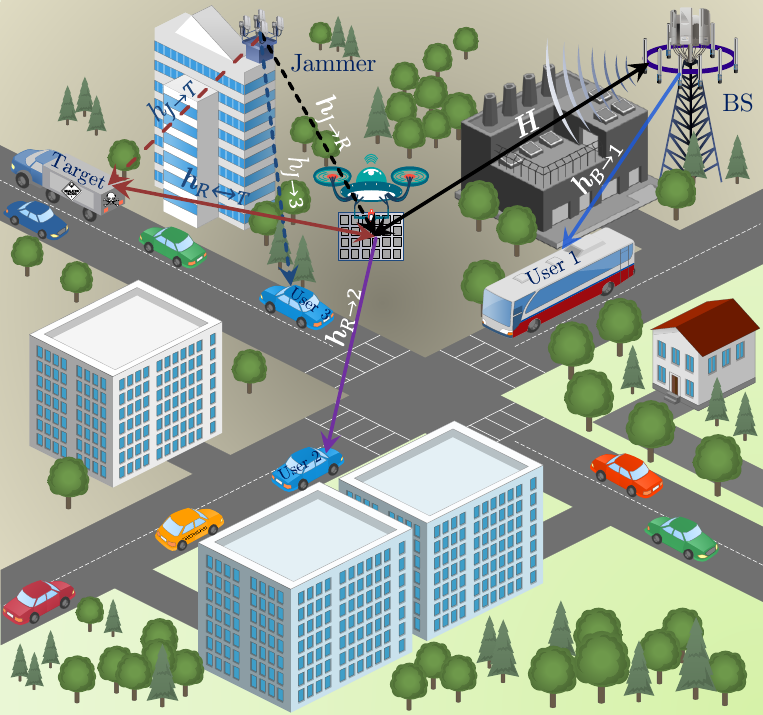}
\caption{ISAC in the RHS-aided UAV network with a dual-function BS.}
\label{fig_sys}
\end{figure}

\section{System Model and Problem Formulation}\label{section_2}
\vspace{-0.24em}
We consider an RHS-aided UAV ISAC system with a dual-function BS that utilizes $N_{t}$ antennas to serve $K$ single-antenna users and a specific target, as shown in Fig.~\ref{fig_sys}. Additionally, an operational single-antenna jammer that emits a jamming signal at power $g_{_{\rm{Jam}}}$ exists and is designed to interfere with ISAC transmissions. 
The total flight duration $T$ of the UAV is segmented into $N$ time intervals, each with a duration of $\Delta_t = \frac{T}{N}$. 
These time intervals are considered to be small enough so that the UAV's position can be treated as nearly constant within each interval, which aids in the design of the trajectory and beamforming for ISAC.
It is posited that the BS can determine the target's rough location via echo signals following extensive scanning, although the direct connection from the BS to the target is obstructed by buildings. 
An RHS-aided UAV with $M$ reflective elements strategically establishes a LoS for target detection and helps users counteract the jamming. We define $\mathbf{\Theta}[n] = \operatorname{diag}[e^{j\theta_1[n]}, \ldots, e^{j\theta_M}]$ as the diagonal RHS phase-shift matrix, where $\theta_m[n] \in [0, 2\pi)$ and $m \in \mathcal{M}=\{1, 2, \ldots, M\}$. 
Moreover, the RHS-aided UAV is designed to remain stable in motion, with its exact ground horizontal coordinates noted as 
$\mathcal{Q}[n] = [x_R[n], y_R[n]]$ when it is stationary in the air. 
Furthermore, we define the UAV's 2-D horizontal velocity in time slot $n$ as $\mathbf{v}[n] \triangleq (v_{R}^x[n], v_{R}^y[n])$. 
For effective signal reception, RHS-aided UAV must maintain a position within a defined region of 
$\|\mathcal{Q}[n] \|_2 \leq \mathbf{r}_0[n]$,
where $\mathbf{r}_0[n]$ is the radius of the circular service area. 
For simplicity, we consider the BS, jammer, target, and users to be stationary on the ground with their respective coordinates designated as $s_{\digamma}[n] = [x_{\digamma}[n], y_{\digamma}[n]]$, where $\digamma \in \{B, R, J, T, k\}$ and $k \in \mathcal{K}=\{1, 2, \ldots, K\}$. 
Assuming that the UAV maintains a constant altitude of $L$, which is above the height of the tallest obstacles within the service area, the distance between the BS or jammer and the RHS becomes 
$\delta_{\Gamma,R}[n] = \sqrt{\|s_{\Gamma}[n] - s_R[n]\|^2 + L^2}, \Gamma \in \{B,J\}$. 
Consequently, the communication channel linking the RHS-aided UAV to either the BS or the jammer can be described accordingly:
\begin{align}
\mathbf{H}[n] &= \sqrt{\frac{{\rm{PL}}_{\alpha}}{\left(\delta_{B,R}[n]\right)^{\beta}}}  \widetilde{\mathbf{H}}[n], \\
\mathbf{h}_{J,R}[n] &= \sqrt{\frac{{\rm{PL}}_{\alpha}}{\left(\delta_{J,R}[n]\right)^{\beta}}} \mathbf{\widetilde h}_{J,R}[n]. 
\end{align}
In this model, ${\rm{PL}}_{\alpha}$ denotes the initial power gain of the channel when measured at a baseline distance of 1 meter, while $\beta$ serves as the path-loss exponent. 
The channel matrix $\widetilde{\mathbf{H}}[n] \in \mathbb{C}^{ M \times N_t}$ is then defined as follows:
\begin{align}
\widetilde{\mathbf{H}}[n] &= \left[ 1, e^{-j \frac{2\pi d_r \cos(\zeta[n])}{\lambda_{{\rm{c}}}}}, \cdots, e^{-j \frac{2\pi d_r (M-1) \cos(\zeta[n])}{\lambda_{{\rm{c}}}}} \right]^H \nonumber
\\ 
&\otimes \left[ 1, e^{-j \frac{2\pi d_t \sin(\xi[n])}{\lambda_{{\rm{c}}}}}, \cdots, e^{-j \frac{2\pi d_t (N_t-1) \sin(\xi)[n]}{\lambda_{{\rm{c}}}}} \right], 
\end{align}
where $d_r$ specifies the uniform spacing between the elements of the RHS, while $d_t$ measures the separation between consecutive antennas at the BS, and $\lambda_{{\rm{c}}}$ is the wavelength of the carrier. Furthermore, the vertical and horizontal angles of arrival (AoAs) at the RHS-aided UAV are denoted by $\zeta[n]= \arccos{\frac{|x_B[n] - x_R[n]|}{\delta_{B,R}[n]}}$ and $\xi[n] = \arcsin{\frac{|y_B[n] - y_R[n]|}{\delta_{B,R}[n]}}$, respectively.
Likewise, $\mathbf{h}_{J,R}[n] \in \mathbb{C}^{M \times 1}$ is represented as:
\begin{align}
\mathbf{\widetilde{h}}_{J,R}[n] &= 
e^{-j \frac{2\pi \delta_{J,R}[n]}{\lambda_{{\rm{c}}}}} 
\\&\times \nonumber
\left[ 1, e^{-j \frac{2\pi d_r \cos(\omega[n])}{\lambda_{{\rm{c}}}}}, \cdots, e^{-j \frac{2\pi d_r (M-1) \cos(\omega[n])}{\lambda_{{\rm{c}}}}} \right]\!, 
\end{align}
where $\omega[n]=\arccos{\frac{|x_J[n] - x_R[n]|}{\delta_{J,R}[n]}}$ represents the AoA from the jammer to the RHS.
The communication link between the RHS and the $k-$th user is expressed as:
\begin{equation}
\mathbf{h}_{R,k}[n] = \sqrt{{\rm{PL}}_{\alpha} \left(\delta_{R,k}[n]\right)^{-\beta}} \mathbf{\widetilde{h}}_{R,k}[n].
\end{equation}
where $\delta_{R,k}$ is the distance between the RHS and the $k$-th user and $\mathbf{\widetilde{h}}_{R,k}[n] \in \mathbb{C}^{M \times 1}$ is defined by:
\begin{equation}
\mathbf{\widetilde{h}}_{R,k}[n] = \left[ 1, e^{-j \frac{2\pi d_r \cos(\phi_k[n])}{\lambda_{{\rm{c}}}}}, \cdots, e^{-j \frac{2\pi d_r (M-1) \cos(\phi_k[n])}{\lambda_{{\rm{c}}}}} \right], 
\end{equation}
where $\phi_k[n] = \arccos{\frac{|x_R[n] - x_{k}[n]|}{\delta_{R,k}[n]}}$ describes the cosine of the angle of departure (AoD) from the RHS to the $k$-th user.

The channel dynamics between the BS and the jammer to ground users typically exhibit Rician fading characteristics. Consequently, the channel vector $\mathbf{h}_{B,k}[n]$ from the BS to the $k$-th user is characterized by
\begin{align}
\mathbf{h}_{B,k}[n] 
&= \sqrt{\frac{{\rm{PL}}_{\alpha}}{\left(\delta_{B,k}[n]\right)^{\beta}}} \bigg(\sqrt{\frac{\varrho}{\varrho + 1}} \mathbf{h}_{B,k}[n]^{(\text{LoS})}[n] 
\\ \nonumber
&+ \sqrt{\frac{1}{\varrho + 1}} \mathbf{h}_{B,k}^{(\text{NLoS})}[n]\bigg) = \sqrt{{\rm{PL}}_{\alpha} \left(\delta_{B,k}[n]\right)^{-\beta}} \widetilde{\mathbf{h}}_{B,k}[n], 
\end{align}
where 
$\varrho$ denotes the Rician factor, and $\delta_{B,k}[n]$ measures the distance between the BS and the $k$-th user.

In a similar way, the channel gains from the jammer to the $k$-th user and the target can be written as:
\begin{align}
h_{J,\beth}[n] &= \sqrt{\frac{{\rm{PL}}_{\alpha}}{\left(\delta_{J,\beth}[n]\right)^{\beta}}} \left(\!\!\sqrt{\frac{\varrho}{\varrho \!+\! 1}} h_{J,\beth}^{(\text{LoS})}[n] \!+\! \sqrt{\frac{1}{\varrho \!+\! 1}} h_{J,\beth}^{(\text{NLoS})}[n]\!\right) \nonumber \\
&=\! \sqrt{{\rm{PL}}_{\alpha} \left(\delta_{J,\beth[n]}\right)^{-\beta}} \widetilde{h}_{J,\beth}[n],
\end{align}
where the distance $\delta_{J,\beth}[n] = \|s_J[n] - s_{\beth}[n]\|, \beth \in \{T,k\}$.

{Under the assumptions that channels are subject to quasi-static flat fading and that perfect channel state information (CSI) is available, the transmission signal is formulated as:
\begin{equation}
\mathbf{x}_B[n] = \sum_{k\in \mathcal{K}} \mathbf{w}_k[n] s_k[n], 
\end{equation}
where $s_k[n] \sim \mathcal{CN}(0,1)$ represents the secure message intended for the $k$-th user, and $\mathbf{w}_k[n] \in \mathbb{C}^{N_{t} \times 1}$ is the beamforming vector for transmitting to this user. The signal received by the $k$-th user is then:
\begin{align}
y_k[n] &= \left(\mathbf{h}_{B,k}^H[n] + \mathbf{h}_{R,k}^H[n] \mathbf{\Theta}[n] \mathbf{H}[n]\right) \mathbf{x}_B[n]  
\nonumber\\
&+\left(h_{J,k} + \mathbf{h}_{R,k}^H[n] \mathbf{\Theta}[n] \mathbf{h}_{J,R}[n]\right) g_{_{\rm{Jam}}} + z_0[n], 
\end{align}
where $z_0[n] \sim \mathcal{CN}(0,\sigma^2)$ is the additive white Gaussian noise (AWGN). 
The composite channel vectors from the BS and jammer to the $k$-th user are denoted 
$\widehat{\mathbf{h}}_{B,k}[n] = \mathbf{h}^{H}_{B,k}[n] + \mathbf{h}^{H}_{R,k}[n] \mathbf{\Theta}[n] \mathbf{H}[n]$ 
and 
$\widehat{{h}}_{J,k}[n] = {h}_{J,k}[n] + \mathbf{h}^{H}_{R,k}[n] \mathbf{\Theta}[n] \mathbf{h}_{J,R}[n]$, respectively. The resulting SINR at the $k$-th user's receiver is defined by:
\begin{equation}
\gamma_{\operatorname{comm},k}[n] \!=\! \frac{\left| \widehat{\mathbf{h}}_{B,k}[n] \mathbf{w}_k[n] \right|^2}{\sum_{i \in \mathcal{K}\backslash k} \left| \widehat{\mathbf{h}}_{B,k}[n] \mathbf{w}_i[n] \right|^2 \!\!+\! g_{_{\rm{Jam}}} \left| \widehat{h}_{J,k}[n] \right|^2 \!\!+\! \sigma^2}. 
\end{equation}
To detect the obstructed target efficiently, a virtual LoS path can be created using the RHS-aided UAV. The sensing performance is evaluated by the beampattern gain. The signal reflected at the RHS is described by
\begin{equation}
\mathbf{x}_R[n] = \mathbf{\Theta}[n] \mathbf{H}[n] \left( \sum_{k\in \mathcal{K}} \mathbf{w}_k[n] s_k[n] \right). 
\end{equation}
The beampattern gain of the RHS in the direction of angle $\varsigma[n]$ is defined as:
\begin{align}
g_{\rm{beam}}(\varsigma[n]) &= 
\mathbf{a}^H(\varsigma[n]) \mathbf{\Theta}[n] \mathbf{H}[n] \left(\sum_{k\in \mathcal{K}} \mathbf{w}_k[n] \mathbf{w}_k^H[n] \right) 
\nonumber\\&\times
\mathbf{H}^H[n] \mathbf{\Theta}[n]^H \mathbf{a}(\varsigma[n]), \vspace*{3mm}
\end{align}
where 
$\mathbf{a}(\varsigma[n]) = 
[1, e^{j2\pi \rho \sin(\varsigma[n])}, 
\ldots, 
e^{j2\pi (M-1) \rho \sin(\varsigma[n])}]^T$ 
represents the steering vector at the RHS, and $\rho$ denotes the distance between adjacent RHS elements, normalized by the wavelength.
Thus, the echo SINR for the target is given by:
\vspace{3mm}\begin{equation}\vspace{3mm}
\gamma_{\operatorname{echo}}[n] = \frac{g_{\rm{beam}}(\varsigma[n])}{g_{_{\rm{Jam}}} \left| h_{J,T}[n] \!+\! \mathbf{a}(\varsigma[n]) \mathbf{\Theta}[n] \mathbf{h}_{J,R}[n] \right|^2 \!+\! \sigma^2}. 
\end{equation}
We assume that the UAV maintains a constant cruising speed during each time slot. 
Based on classic aerodynamic theory for rotary-wing UAVs, the aerodynamic power consumption for level flight during time slot $n$ can be modeled as:
\begin{equation} 
P_{\mathrm {aero}}[n]=
P_{\mathrm {profile}}[n]+
P_{\mathrm {parasite}}[n]+
P_{\mathrm {induced}}[n],
\end{equation}
where $P_{\text{profile}}[n]$, $P_{\text{parasite}}[n]$, and $P_{\text{induced}}[n]$ represent the blade profile power, parasite power, and induced power, respectively, and are given by \cite{seddon2011basic}:
\begin{align} 
P_{\mathrm {profile}}[n]
&=P_{0}\left(1 + \frac{3 \| \mathbf{v}[n] \|^{2}}{\Omega ^{2} r^{2}}\right)  \\ 
P_{\mathrm {parasite}}[n]
&= \frac{1}{2} \b{o} \mho \ss \AA \| \mathbf{v}[n] \|^{3} \\ 
P_{\mathrm {induced}}[n]
&= P_{I} \left(\sqrt{1 + \frac{\| \mathbf{v}[n] \|^{4}}{4 v_{0}^{4}}} - \frac{\| \mathbf{v}[n] \|^{2}}{2 v_{0}^{2}}\right)^{\frac{1}{2}},
\end{align}
where $P_0$ and $P_I$ refer to the inherent power of the blade profile and the induced power when $v[n] = 0$, respectively. 
The symbols $\Omega$ and $r$ represent the angular speed of the blade and the radius of the rotor. The parameters $\b{o}$ and $\mho$ stand for the drag coefficient of the fuselage and the density of the air, respectively. The rotor solidity is denoted by $\ss$, the rotor disc area by $\AA$, and the average induced velocity by the rotor is indicated as $v_0$~\cite{seddon2011basic,8663615,9447216}.


To counteract jamming disruptions in ISAC systems, our objective is to reduce the total network power by collectively adjusting active and passive beamforming techniques, and positioning of the RHS-aided UAV, ensuring the SINR meets the necessary criteria for target detection and communication users' QoS. Thus, we need to solve the MINLP:
\begin{subequations}
\label{main_op}
\begin{align}
&~\text{P}_1:
\min_{\mathbf{w}_k[n], \mathbf{\Theta}[n], \mathcal{Q}[n],\mathbf{v}[n]}
\eta \sum_{k\in \mathcal{K}}  \left|\mathbf{w}_k[n]\right|^2 
+ P_{\operatorname{aero}}[n] 
\nonumber\\
&~\quad\quad\quad\quad\quad\quad \quad\quad\quad  
+ (N_t+1)P_{\operatorname{circ}}
+ g_{_{\rm{Jam}}} 
\\
&s.t.: ~ \sum\nolimits_{k\in \mathcal{K}} \left|\mathbf{w}_k[n]\right|^2 \leq g_{{\rm{BS}},\max}, 
\label{p1_c1}\\
&~\quad\quad \sum\nolimits_{k\in \mathcal{K}} \log_2(1+ \gamma_{\operatorname{comm},k}[n]) \geq R_{\min},  
\label{p1_c2}\\
&~\quad\quad \gamma_{\operatorname{echo}}[n] \geq \gamma_T,  
\label{p1_c3}\\
&~\quad\quad \|\mathcal{Q}[n] \|_2 \leq \mathbf{r}_0[n], 
\label{p1_c4}\\
&~\quad\quad \|\mathcal{Q}[n+1] - \mathcal{Q}[n]\|_2 \leq \mathbf{v}[n]\Delta_t, 
\label{p1_c5}\\
&~\quad\quad \|\mathbf{v}[n+1] - \mathbf{v}[n]\|_2 \leq a_{\max}\Delta_t, 
\label{p1_c6}\\
&~\quad\quad |\mathbf{v}[n]| \leq V_{\max}, 
\label{p1_c7}\\
&~\quad\quad \left|\theta_m[n] \right| = 1, \forall m \in \mathcal{M}. 
\label{p1_c8}
\end{align}
\end{subequations}
where $\eta > 1$ and $P_{\operatorname{circ}}$ represent the power amplifier efficiency and the circuit power consumption of a single antenna's RF chain, respectively. 
Constraint \eqref{p1_c1} sets a limit on the transmit power of the BS to $g_{{\rm{BS}},\max}$, which is determined by the analog RF front-end. 
The parameter $R_{\min}$ in \eqref{p1_c2} represents the minimum data rate required by user $k$ to meet the QoS requirements, while $\gamma_T$ in \eqref{p1_c3} is the defined minimum echo SINR threshold for target detection. 
\eqref{p1_c4} specifies the radius of the circular service area of the RHS-aided UAV.
Constraint \eqref{p1_c5} models the UAV's trajectory evolution based on its flight velocity.
Constraint \eqref{p1_c6} limits the change in the UAV's speed from one time slot to the next, where $a_{\text{max}}$ is the maximum acceleration the UAV can achieve, restricted by its engines, and \eqref{p1_c7} limit the UAV's maximum velocity to $v_{\text{max}}$. 
Finally, constraint \eqref{p1_c8} stands for the unit modules constraint. 
Clearly, the optimization problem \eqref{main_op} is non-convex due to the intertwined nature of the variables, non-convex objective, and constraints.
We note that since $(N_t+1) \cdot P_{\operatorname{circ}}+g_{_{\rm{Jam}}}$ is a constant for a given number of antenna elements, it is omitted in the solution of \eqref{main_op}.

\section{Proposed Iterative RHOSI Algorithm}\label{section_3}
To solve $\text{P}_1$, we introduce an AO algorithm designed to enhance transmit beamforming, adjust phase shifts, and strategically position the RHS-aided UAV to effectively counteract jamming threats, while adhering to the echo SINR specifications for target detection. Initially, active beamforming parameters are determined using successive convex approximation (SCA). Subsequently, passive beamforming is refined through a penalty-based optimization technique. The placement of the RHS-aided UAV is determined using SCA to secure a viable configuration. These components are iteratively refined to achieve optimal performance~\cite{10872813}.

This integrated design framework, referred to as RHOSI, minimizes the total network power in a unified loop. Unlike conventional methods that address each optimization block in isolation, RHOSI explicitly leverages their interdependence: transmit beams adapt to user and jammer dynamics, RHS phases reshape the propagation environment by steering constructive paths and suppressing interference, and UAV positioning ensures favorable geometries for both communication and sensing. By alternating across these dimensions, RHOSI converges to a feasible and robust solution that strengthens throughput while safeguarding sensing reliability under jamming.
The main steps of RHOSI are shown in \textbf{Algorithm~\ref{alg_final}}.
\begin{algorithm}[t]
\caption{\strut Proposed RHOSI Algorithm}
\begin{algorithmic}[1]\label{alg_final}
\renewcommand{\algorithmicrequire}{\textbf{Input:}}
\renewcommand{\algorithmicensure}{\textbf{Output:}}
\REQUIRE Set $s=0$, set maximum number of iteration $\mathcal{S}_{\max}$,
initialize
$\mathbf{W}_k[n]=\mathbf{W}_k^{(0)}[n]$,
$\mathbf{\Theta}[n]=\mathbf{\Theta}^{(0)}[n]$, and
$\mathcal{Q}[n]=\mathcal{Q}^{(0)}[n],$
$\mathbf{v}^{}[n]=\mathbf{v}^{(\mathrm{0})}[n]$.\\    
\STATE \textbf{repeat}\\
\STATE \quad Solve $\text{P}_1$
to obtain the optimal solution $\mathbf{W}_k^{(s)}[n]$ for \\ \quad given 
$\mathbf{\Theta}^{(s-1)}[n]$ and
 $(\mathcal{Q}^{(s-1)}[n],\mathbf{v}^{(s-1)}[n])$.
\STATE \quad Solve $\text{P}_1$
to obtain the optimal solution $\mathbf{\Theta}^{(s)}[n]$ for \\ \quad given 
$\mathbf{W}_k^{(s-1)}[n]$ and
$(\mathcal{Q}^{(s-1)}[n],\mathbf{v}^{(s-1)}[n])$ .
\STATE \quad Solve 
$\text{P}_1$
for given 
$\mathbf{W}_k^{(s-1)}[n]$ and
$\mathbf{\Theta}^{(s-1)}[n]$ to obtain \\ \quad the optimal solution $(\mathcal{Q}^{(s)}[n],\mathbf{v}^{(s)}[n])$.
\STATE   \textbf{until} ${s}=\mathcal{S}_{\max}$
\STATE   \textbf{return} 
$\{
\mathbf{W}_k^{(s)}[n],\mathbf{\Theta}^{(s)}[n],(\mathcal{Q}^{(s)}[n],\mathbf{v}^{(s)}[n])
\}=$ 
$\{
\mathbf{W}_k^{(\mathrm{opt})}[n],
\mathbf{\Theta}^{(\mathrm{opt})}[n],
(\mathcal{Q}^{(\mathrm{opt})}[n],\mathbf{v}^{(\mathrm{opt})}[n])
\}$
\end{algorithmic}
\end{algorithm}	

\vspace{-7mm}
\section{Simulation Results}\label{section_5}
\vspace{-2mm}
In this section, we assess the performance of the proposed RHOSI algorithm via computer simulations. The deployment area is a $0.4 \text{ km} \times 0.4 \text{ km}$ square area, where the positions of the BS, jammer, and target are assumed fixed. The BS is equipped with $N_{t}=6$ antennas and serves $K=3$ users. The minimum long-term sensing\slash echo SINR in the UAV is $\gamma_T=3 \text{ dB}$. The UAV's maximum flight speed with $M=20$ RHS elements is $v_{\text{max}}=15 \text{ m/s}$, and its flight altitude is $L=40 \text{ m}$. The channel power gain at the reference distance $d_0=1 \text{ m}$ is ${\rm{PL}}_{\alpha}=-20 \text{ dB}$. Unless stated otherwise, we set $\sigma^2=-90 \text{ dBm}$, $g_{{\rm{BS}},\max}=40 \text{ dBm}$, $g_{_{\rm{Jam}}}=30 \text{ dBm}$, $R_{\text{min}}=1 \text{ bps/Hz}$, $a_{\text{max}}=5 \text{ m/s}^2$, $T=60 \text{ s}$, and $\Delta_t=1 \text{ s}$~\cite{9858656}. 

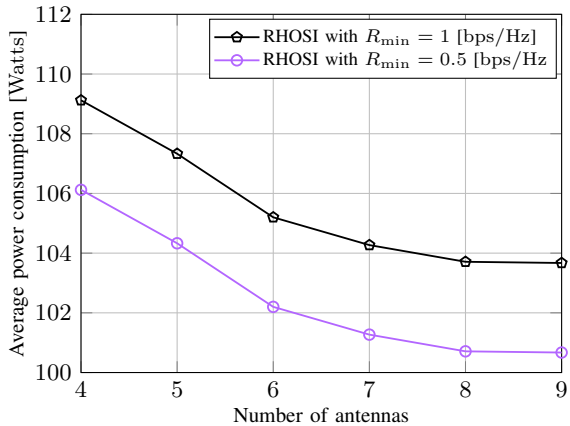
\begin{figure}[t]
\centering
\begin{tikzpicture}
\begin{axis}[
    width=0.90\columnwidth,
    height=0.35\textwidth,
    xlabel={Number of antennas},
    ylabel={Average power consumption [$\rm{Watts}$]},
    grid=major,
    legend style={
        at={(0.99,0.82)}, 
        anchor=south east, 
        font=\scriptsize,
        inner sep=0.5mm,
        legend cell align={left},
        legend columns=1,
        /tikz/column 1/.style={column sep=-1pt,},
        /tikz/column 2/.style={column sep=0pt,},
        /tikz/row 1/.style={row sep=-2pt,},
        /tikz/row 2/.style={row sep=-2pt,},
        /tikz/row 3/.style={row sep=-5pt,},
    },
    tick label style={font=\small},
    xlabel style={font=\footnotesize, yshift=1.00mm},
    ylabel style={font=\footnotesize, yshift=-1.5mm},
    xmin=4, xmax=9,
    xtick={4,5,6,7,8,9},
    ymin=100, ymax=112,
    ytick={100,102,104,106,108,110,112},
    legend entries={
        RHOSI with $R_{\min}=1$ [$\rm{bps/Hz}$],
        RHOSI with $R_{\min}=0.5$ [$\rm{bps/Hz}$,
        $~~~~$Method B,
        Method C,
        $~~~~$Method D,
        Method E,
        $~~~~~~$Initial
    }
]

\addplot[mark=pentagon, solid, black, line width=0.7pt] coordinates {
    (4, 109.12) (5, 107.33) (6, 105.2) (7, 104.27) (8, 103.71) (9, 103.67)
};
\addplot[mark=o, solid, lightpurple, line width=0.7pt] coordinates {
    (4, 106.12) (5, 104.33) (6, 102.2) (7, 101.27) (8, 100.71) (9, 100.67)
};

\end{axis}
\end{tikzpicture}
\caption{Average power versus the number of antennas.}
\label{fig2}
\end{figure}
Fig.~\ref{fig2} illustrates the average power consumption as a function of the number of antennas at the BS under different minimum data rate requirements $R_{\min}$. It can be observed that RHOSI achieves a clear reduction in power consumption as the number of BS antennas increases. This trend is explained by the additional spatial degrees of freedom (DoFs) offered by larger BS antenna arrays, which enable more precise beamforming and interference mitigation. Nevertheless, the rate of improvement reduces as the number of antennas grows, reflecting the diminishing returns of adding further antennas. A comparison between the curves shows that stricter QoS requirements (higher $R_{\min}$) consistently result in higher power consumption. This is because meeting a higher minimum rate forces the BS to transmit with stronger beams and maintain tighter interference suppression, increasing energy usage. 

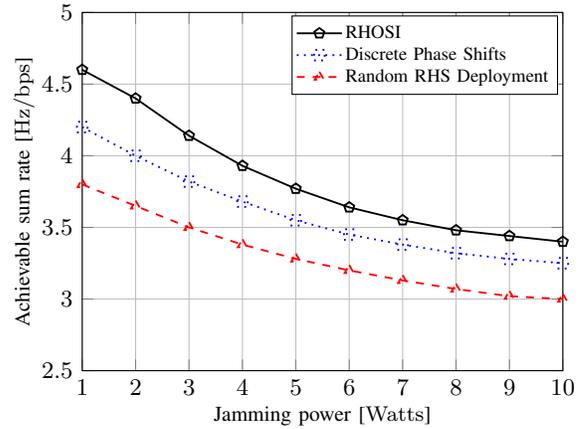
\begin{figure}[t]
\centering
\begin{tikzpicture}
\begin{axis}[
    width=0.90\columnwidth,
    height=0.35\textwidth,
    xlabel={Jamming power [$\rm{Watts}$]},
    ylabel={Achievable sum rate [$\rm{Hz/bps}$]},
    grid=major,
    legend style={
        at={(0.99,0.77)}, 
        anchor=south east, 
        font=\scriptsize,
        inner sep=0.5mm,
        legend cell align={left},
        legend columns=1,
        /tikz/column 1/.style={column sep=-1pt,},
        /tikz/column 2/.style={column sep=0pt,},
        /tikz/row 1/.style={row sep=-2pt,},
        /tikz/row 2/.style={row sep=-2pt,},
        /tikz/row 3/.style={row sep=-5pt,},
    },
    tick label style={font=\small},
    xlabel style={font=\footnotesize, yshift=1.00mm},
    ylabel style={font=\footnotesize, yshift=-1.5mm},
    xmin=1, xmax=10,
    xtick={1,2,3,4,5,6,7,8,9,10},
    ymin=2.5, ymax=5,
    ytick={2.5,3,3.5,4,4.5,5},
    legend entries={
        RHOSI,
        Discrete Phase Shifts,
        Random RHS Deployment,
    }
]

\addplot[mark=pentagon, solid, black, line width=0.7pt] coordinates {
    (1, 4.6) 
    (2, 4.4) 
    (3, 4.14) 
    (4, 3.93) 
    (5, 3.77) 
    (6, 3.64) 
    (7, 3.55) 
    (8, 3.48) 
    (9, 3.44) 
    (10, 3.40)
};

\addplot[mark=square, dotted, blue, line width=0.7pt] coordinates {
    (1, 4.2) 
    (2, 4.0) 
    (3, 3.82) 
    (4, 3.68) 
    (5, 3.55) 
    (6, 3.45) 
    (7, 3.38) 
    (8, 3.32) 
    (9, 3.28) 
    (10, 3.25)
};
\addplot[mark=triangle, dashed, red, line width=0.7pt] coordinates {
    (1, 3.8) 
    (2, 3.65) 
    (3, 3.5) 
    (4, 3.38) 
    (5, 3.28) 
    (6, 3.20) 
    (7, 3.13) 
    (8, 3.07) 
    (9, 3.02) 
    (10, 3.0)
};
\end{axis}
\end{tikzpicture}
\caption{The achievable sum rate versus the transmit power
of the jammer.}
\label{fig3}
\end{figure}

Fig.~\ref{fig3} depicts the achievable sum-rate versus the jammer power $g_{_{\rm{Jam}}}$. As expected, all schemes degrade as $g_{_{\rm{Jam}}}$ increases. Notably, the slope of degradation flattens at higher $g_{_{\rm{Jam}}}$, indicating that the passive beamforming gain from the RHS becomes increasingly valuable under strong jamming. By providing extra spatial DoF and diversity, the RHS helps preserve useful signal components and suppress interference, slowing the rate loss even in the high–$g_{_{\rm{Jam}}}$ regime.
At all jamming levels, RHOSI attains the highest sum-rate, while the random RHS deployment case performs the worst. This ordering is intuitive: RHOSI jointly optimizes BS transmit beamforming and RHS phase configuration to coherently combine desired signals toward users and the sensing target, boosting the anti-jamming SINR. In contrast, the discrete phase-shift scheme benefits from structured phase control but suffers a loss due to its limited phase resolution, which restricts the achievable passive beamforming gain. 
Finally, the random RHS deployment scheme performs below the discrete phase-shift baseline, indicating that while an RHS can still introduce extra propagation paths and spatial diversity, non-optimized placement reduces the ability to harness these gains for interference suppression and coherent enhancement.

\vspace{-2mm}
\section{Conclusion and Future Work}\label{section_6}
\vspace{-2mm}
This paper explored joint resource allocation and trajectory planning for an RHS-aided UAV ISAC system. We framed an optimization problem aimed at minimizing the total network's power consumption while meeting the QoS requirements for both communication and sensing users.
The RHOSI algorithm was developed to tackle the non-convex MINLP problem, yielding a high-quality suboptimal solution. Simulations demonstrated significant power savings with RHOSI as compared to baseline methods.

\bibliographystyle{ieeetr}
\bibliography{ref}
\end{document}